\documentclass[aps,pra,reprint,showpacks,onecolumn,notitlepage,superscriptaddress]{revtex4-1}

\usepackage{amsfonts}
\usepackage{mathrsfs}
\usepackage{amsmath}
\usepackage{amssymb}
\usepackage{graphicx}
\usepackage{dcolumn}
\usepackage{color}%
\usepackage{bm}
\usepackage{soul}
\usepackage{mathtools}
\usepackage{booktabs}
\newcommand{\figpanel}[2]{\hyperref[#1]{\ref*{#1}(#2)}}

\usepackage[colorlinks,linkcolor=blue,citecolor=blue,hyperindex,bookmarks=false,pdfstartview=FitH]{hyperref}

\begin{document}

\title[Giant atoms in a structured bath with broken time-reversal symmetry]{Decay dynamics of a giant atom in a structured bath with broken time-reversal symmetry}

\author{Lei Du}
\affiliation{Center for Quantum Sciences and School of Physics, Northeast Normal University, Changchun 130024, China}
\affiliation{Center for Theoretical Physics and School of Science, Hainan University, Haikou 570228, China}
\author{Yao-Tong Chen} 
\affiliation{Center for Quantum Sciences and School of Physics, Northeast Normal University, Changchun 130024, China}
\author{Yan Zhang}
\affiliation{Center for Quantum Sciences and School of Physics, Northeast Normal University, Changchun 130024, China}
\author{Yong Li}
\affiliation{Center for Theoretical Physics and School of Science, Hainan University, Haikou 570228, China}
\author{Jin-Hui Wu}
\email{jhwu@nenu.edu.cn}
\affiliation{Center for Quantum Sciences and School of Physics, Northeast Normal University, Changchun 130024, China}

\date{\today}

\vspace{10pt}

\begin{abstract}
We study in this paper the decay dynamics of a two-level giant atom, which is coupled to a quasi-one-dimensional sawtooth lattice exposed to uniform synthetic magnetic fluxes. In the case where the two sublattices have a large detuning, the giant atom is effectively coupled to a single-band structured bath with flux-controlled energy band and time-reversal symmetry. This feature significantly affects the decay dynamics of the giant atom as well as the propagation of the emitted photon. In particular, the giant atom can exhibit chiral spontaneous emission and allow for nonreciprocal delayed light, which are however unattainable by coupling a small atom to this lattice. Giant atoms with different frequencies can be designed to emit photons towards different directions and with different group velocities. Our results pave the way towards engineering quantum networks and manipulating giant-atom interference effects.
\end{abstract}

\maketitle

\section{Introduction}

Giant atoms, which can be understood as quantum emitters coupled to a (bosonic) field at multiple separate points, have emerged as a burgeoning quantum optical paradigm due to their self-interference effects~\cite{fiveyear,LambAFK}. These peculiar quantum interferences can significantly affect the atom-field interactions and thus can be harnessed to engineer various building blocks of quantum networks and quantum many-body systems. Despite the simple architectures of giant atoms, a large number of unconventional phenomena have been witnessed with this paradigm, such as frequency-dependent Lamb shifts and relaxation rates~\cite{LambAFK,GLZ2017,nonexp}, exotic bound states~\cite{WXchiral1,oscillate,ZhaoWbound,VegaPRA,YuanGA,ChengGA,LimGA}, chiral atom-field interactions~\cite{DLprl,WXchiral2,ChenCP}, decoherence-free interactions~\cite{NoriGA,braided,FCdeco,complexDFI}, and non-Markovian scattering spectra~\cite{JiaGA,DLprr1,ZhuEIT}, to name a few.

While most of the existing works concentrated on giant atoms coupled to (one-dimensional) continuous waveguides such as superconducting transmission lines, it is quite interesting and significant to study the interactions between giant atoms and discrete photonic lattices. In contrast to continuous waveguides, photonic lattices can be engineered to feature rich band structures with tailored energy bands and gaps. This thus allows us to design and manipulate photon-atom bound states~\cite{bdfirst,bd1,bd2,bd3,bd4}. Moreover, such structured baths typically exhibit stronger memory effects and thus provide a powerful platform for simulating, e.g., quantum Zeno and anti-Zeno effects~\cite{ZenoAn,ZenoDL}. Some recent works have studied the decay dynamics, bound states, and single-photon scatterings for giant atoms coupled to lattices with time-reversal-symmetric dispersion relations~\cite{ZhaoWbound,YuanGA,LimGA,LonghiGA,AFKstructured,magicQED} and even with topologically nontrivial phases~\cite{VegaPRA,ChengGA,Vega2D}. When considering photonic lattices with periodically arranged closed loops, one can expect richer giant-atom interference effects by breaking the time-reversal symmetry of the bath via synthetic magnetism and creating flat bands~\cite{flatPRA,Flach2014,Flach2016,flatQST,sawtoothAGT}. In a very recent work~\cite{Hofstadter}, Wang \emph{et al.} studied giant atoms coupled to a Hofstadter-ladder waveguide whose time-reversal symmetry can be broken by tuning the synthetic magnetic field threading each square plaquette. However, this work~\cite{Hofstadter} focused on the regime where the atomic frequency lies within the band gap so that only virtual photon processes can occur.

In this paper, we study the decay dynamics of a two-level giant atom coupled to a quasi-one-dimensional sawtooth lattice~\cite{flatPRA,Flach2014,DLjosab}. We assume that the two sublattices are far detuned from each other, so that one can adiabatically eliminate the sublattice with fewer sites and the bath is equivalent to a single-band lattice with modified hopping rates. If each triangle plaquette of the sawtooth lattice is threaded by a synthetic magnetic flux, the effective lattice shows a flux-controlled band structure which collapses into a complete flat band under certain conditions. The dynamical behavior of the giant atom is closely related to the modified band structure, showing, e.g., localized states with oscillating amplitudes and chiral atomic spontaneous emission. We reveal that the chiral emission arises from the direction-dependent giant-atom interference effects, thus it cannot be achieved with a small atom even if the time-reversal symmetry of the bath is broken. The propagation direction and group velocity of the emitted photon can be tailored by properly matching the atomic frequency, the synthetic magnetic flux, and the separation of the coupling points. This provides a global control when considering more giant atoms coupled to this lattice. Moreover, we demonstrate the competition of the chiral effects resulting from the synthetic magnetism of the sawtooth lattice and an additional phase difference between the two atom-field coupling points.

\section{Model and equations}
\begin{figure}[ptb]
\centering
\includegraphics[scale=0.48]{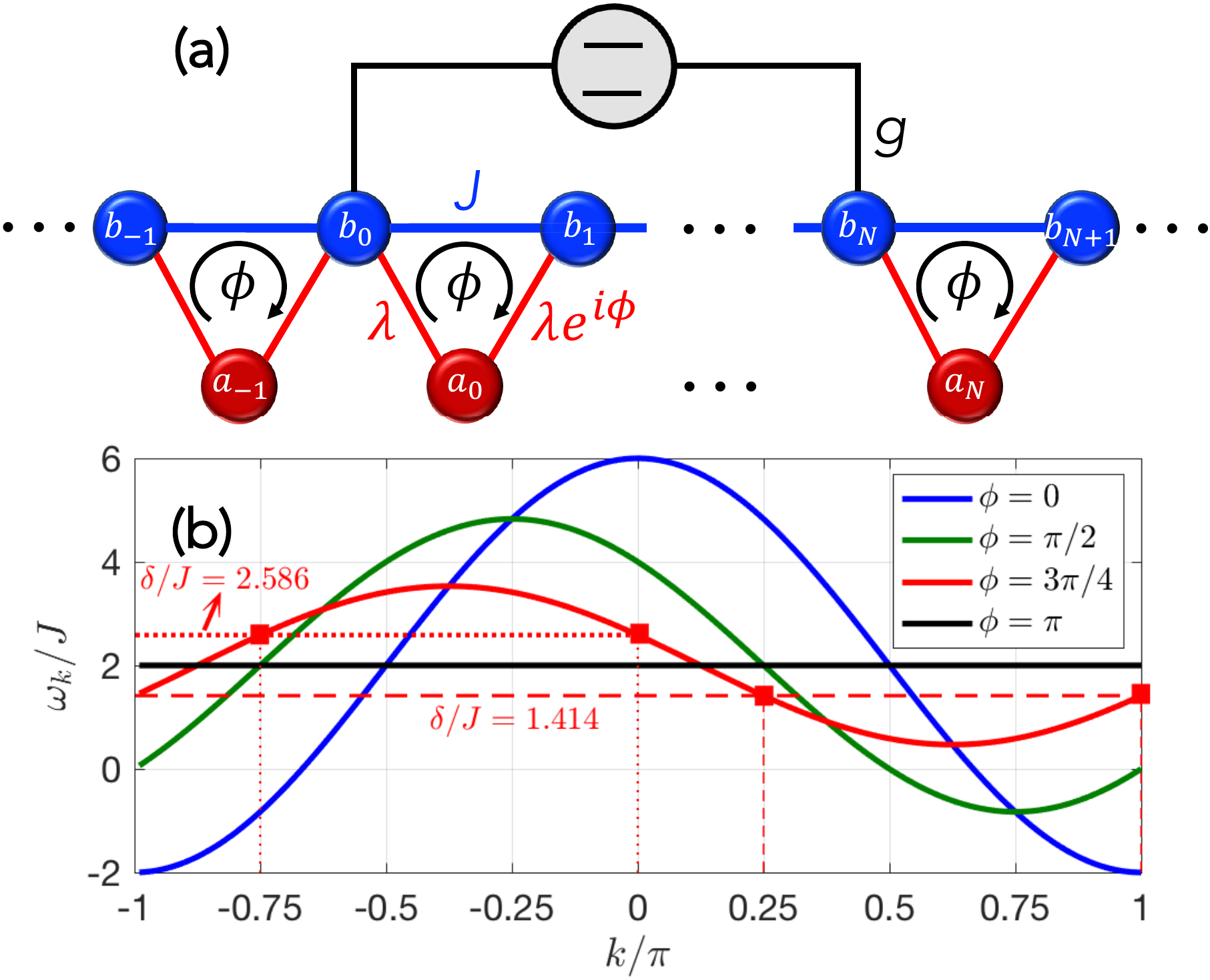}
\caption{(a) Schematic of the giant-atom system containing a sawtooth lattice. (b) Energy band in Eq.~(\ref{dispersion}) as a function of $k$ for $\beta=J$ and different values of the overall phase $\phi$. In panel (b), the red dotted and dashed horizontal lines represent $\delta/J=2.586$ and $\delta/J=1.414$, respectively, and the red squares indicate their intersection points with the energy band of $\phi=3\pi/4$.}
\label{fig1}
\end{figure}

We first consider a quasi-one-dimensional sawtooth lattice comprising two sublattices $A$ and $B$ as shown in Fig.~\figpanel{fig1}{a}, which is described by the Hamiltonian ($\hbar=1$ in this paper)
\begin{eqnarray}
H_{\text{AB}}&=&\sum_{m}\left[(\omega_{0}-\Delta)a_{m}^{\dag}a_{m}+\omega_{0}b_{m}^{\dag}b_{m}\right]+\sum_{m}J\left(b_{m}^{\dag}b_{m+1}+\text{H.c.}\right)\nonumber\\
&&+\sum_{m}\lambda\left[a_{m}^{\dag}(b_{m}+e^{i\phi}b_{m+1})+\text{H.c.}\right],
\label{sawtoothH}
\end{eqnarray}
where $a_{m}$ and $b_{m}$ are the annihilation operators of the $m$th sites of sublattices $A$ and $B$, respectively; $\omega_{0}$ is the resonance frequency of each site of $B$ and $\Delta$ is the frequency detuning between the sites of $A$ and $B$ (both $A$ and $B$ are assumed to be uniform); $\lambda$ describes the coupling between $A$ and $B$ and $J$ is the nearest-neighbor hopping rate of $B$; $\phi$ is the overall coupling phase (synthetic magnetic flux) of each triangle plaquette, which plays a key role in breaking the time-reversal symmetry of the Hermitian Hamiltonian. Such a sawtooth lattice can be experimentally implemented with, e.g., superconducting quantum circuits~\cite{sawtoothAGT,circuit1,circuit2,circuit3,Roushan}.

Now let's consider a two-level giant atom (with ground state $|g\rangle$, excited state $|e\rangle$, and transition frequency $\omega_{a}$) coupled to the sawtooth lattice at two different lattice sites. If the giant atom only interacts with sublattice $B$ and $|\Delta|$ is large compared with $\lambda$, one can adiabatically eliminate sublattice $A$ and the Hamiltonian of our model can be effectively written as (see Appendix~\ref{appa} for more details)
\begin{equation}
H_{\text{eff}}=\delta\sigma_{+}\sigma_{-}+\sum_{m}2\beta b_{m}^{\dag}b_{m}+\sum_{m}\left(\xi b_{m}^{\dag}b_{m+1}+\text{H.c.}\right)+\left[g\sigma_{+}(b_{0}+b_{N})+\text{H.c.}\right],
\label{effectH}
\end{equation}
where $\delta=\omega_{a}-\omega_{0}$ is the frequency detuning between the atom and the sites of $B$ (we have performed a frame rotation with respect to $\omega_{0}$); $\sigma_{-}=(\sigma_{+})^{\dag}=|g\rangle\langle e|$ is the lowering operator of the giant atom; $\beta=\lambda^{2}/\Delta$ represents the effective on-site potential of $B$ arising from the adiabatic elimination of $A$; $\xi=J+\beta\text{exp}(i\phi)$ describes the effective nearest-neighbor hopping of $B$, which does not break the Hermiticity of the Hamiltonian; $g$ is the atom-lattice coupling strength, which is assumed to be identical at the two coupling points $b_{0}$ and $b_{N}$ (we refer to $N$ as the coupling separation hereafter). In the single-excitation subspace, the state of the model at time $t$ can be written as
\begin{equation}
|\psi(t)\rangle=\left[\sum_{m}c_{m}(t)b_{m}^{\dag}+c_{e}(t)\sigma_{+}\right]|g\rangle\otimes|V\rangle,
\label{psi}
\end{equation}
where $c_{e}$ and $c_{m}$ are the excitation amplitudes of the atom and the $m$th site of $B$, respectively, and $|V\rangle$ is the vacuum state of the lattice. In this way, the dynamical equations of the amplitudes can be obtained as
\begin{eqnarray}
\dot{c}_{e}&=&-i\delta c_{e}-ig(c_{0}+c_{N}), \label{de1}\\
\dot{c}_{m}&=&-2i\beta c_{m}-i(\xi c_{m+1}+\xi^{*}c_{m-1})-igc_{e}(\delta_{m,0}+\delta_{m,N}). \label{de2}
\end{eqnarray}
If the atomic transition frequency lies within the energy band of the lattice, the effective decay rate of the giant atom (in the Markovian limit) can be given by~\cite{LambAFK,braided,LonghiGA}
\begin{equation}
\Gamma_{\text{eff}}=4\pi g^{2}\left[1+\cos{(k_{a}N)}\right]D(\omega_{a}),
\label{decayrate}
\end{equation}
where $k_{a}$ and $D(\omega_{a})$ are, respectively, the wave vector and density of states of the field at the atomic frequency. It shows that the giant atom can be made \emph{dissipationless} (other decay channels are neglected) if $k_{a}N$ is an odd multiple of $\pi$. Nevertheless, the decay dynamics of the giant atom can deviate from the prediction of Eq.~(\ref{decayrate}) if the finite non-Markovian retardation effect is taken into account. Moreover, $\Gamma_{\text{eff}}$ becomes ill defined if the effective lattice shows a flat band as will be shown below.

Before studying the dynamics of the giant atom, we would like to demonstrate the energy band of the effective single-band lattice [described by Eq.~(\ref{de2}) with $g=0$], which can be obtained as
\begin{equation}
\omega_{k}=2J\cos{k}+2\beta\left[1+\cos{(k+\phi)}\right]
\label{dispersion}
\end{equation}
by assuming $c_{m}(t)\propto\text{exp}(ikm-i\omega_{k}t)$. Clearly, $\omega_{k}$ is strongly dependent on the value of $\phi$, and in particular, shows a complete \emph{flat band} (i.e., independent of the wave vector $k$) when $J=\beta$ and $\phi=(2n+1)\pi$ ($n$ is an arbitrary integer). In a more general case of $\phi\neq n\pi$, the energy band becomes asymmetric with respect to $k=0$ and the bandwidth changes markedly with $\phi$, as shown in Fig.~\figpanel{fig1}{b}. In view of this, we envision that the giant-atom interference effects, which are closely related to the phase accumulation $\phi=k_{a}N$ of the emitted photon traveling between the two coupling points and its group velocity $v_{g}=\partial\omega_{k}/\partial k$, can be significantly modified by tuning the value of $\phi$.

\section{Flux-controlled giant-atom interference effects}

To study the decay dynamics of the giant atom, we assume that the atomic transition frequency $\omega_{a}$ lies within the energy band of the effective lattice [i.e., $\text{min}(\omega_{k})<\delta<\text{max}(\omega_{k})$, except when the lattice shows a flat band with zero bandwidth]. Moreover, we consider the case of $g\ll\{J,\beta\}$ such that the non-Markovian retardation effect is weak.

\begin{figure}[ptb]
\centering
\includegraphics[scale=0.52]{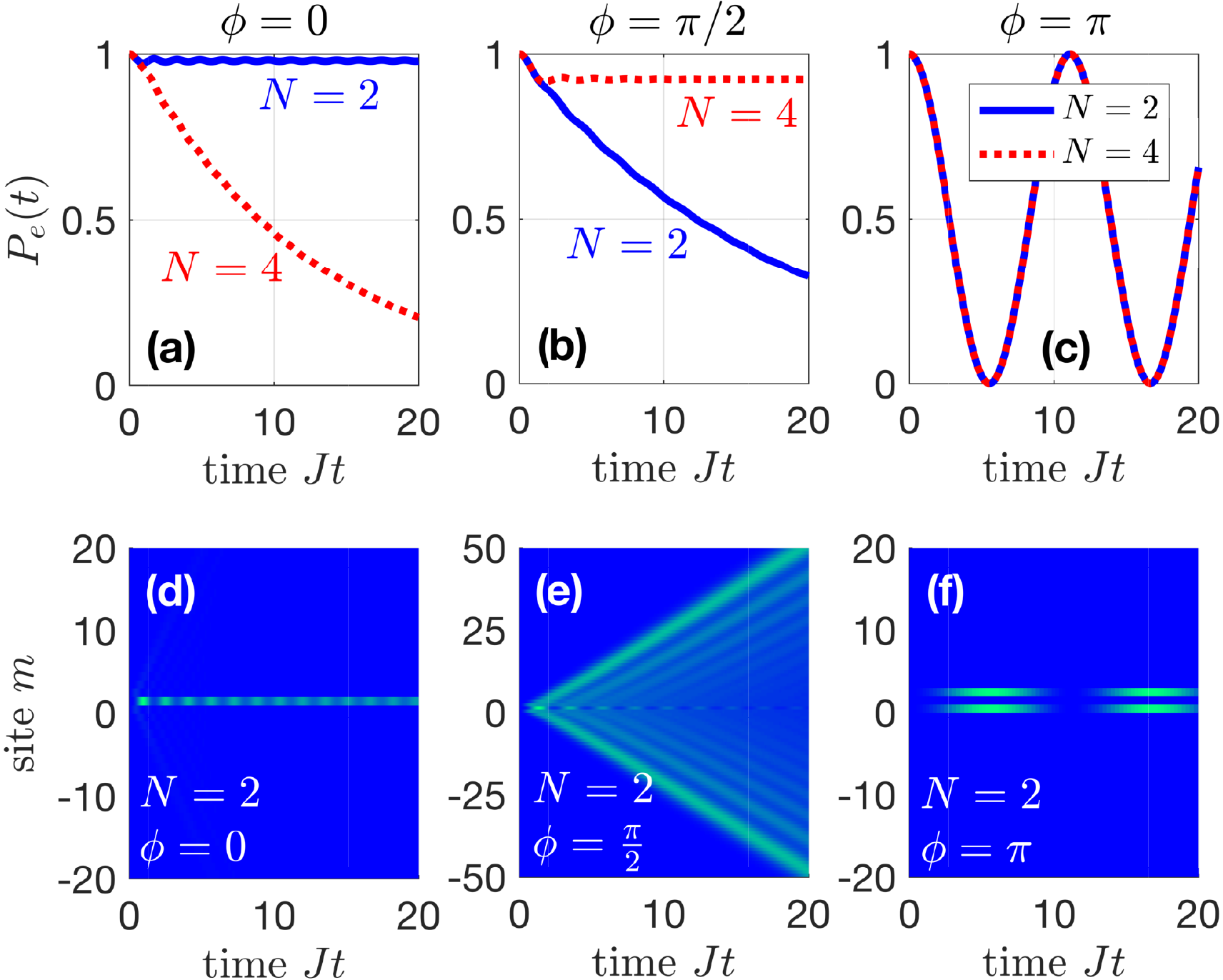}
\caption{(a)-(c) Dynamics of atomic excitation probability $P_{e}(t)$ for different values of coupling separation $N$ and overall phase $\phi$. (d)-(f) Dynamics of lattice probability distribution $P_{m}(t)$ for $N=2$ and different values of $\phi$. Other parameters are $g/J=0.2$, $\beta/J=1$, $\delta=2J$, and $m_{\text{tot}}=200$.}
\label{fig2}
\end{figure}

We first consider the case where the atomic frequency is pinned at the band center of the effective lattice, i.e., $\delta=2\beta$. Figures~\figpanel{fig2}{a}-\figpanel{fig2}{c} show the time evolutions of the atomic excitation probability $P_{e}(t)=|c_{e}(t)|^{2}$ for different values of coupling separation $N$ and overall phase $\phi$ (we assume that the atom is initially in the excited state). We solve Eqs.~(\ref{de1}) and (\ref{de2}) numerically with an effective lattice of $200$ sites in total ($m_{\text{tot}}=200$). When $\phi=2n\pi$, as shown in Fig.~\figpanel{fig2}{a}, the giant atom shows the familiar decay behavior as if it is coupled to a common one-dimensional single-band lattice (with real-valued nearest-neighbor couplings)~\cite{LonghiGA,DLprl,AFKstructured}: the giant atom is nearly dissipationless if $N=2$ and shows a superradiance-like emission if $N=4$. However, this typical behavior can be changed if $\phi\neq 2n\pi$. As shown in Fig.~\figpanel{fig2}{b}, when $\phi=(2n+1/2)\pi$, the giant atom shows a complete decay if $N=2$ and a fractional decay (i.e., the atomic excitation probability does not vanish in the long-time limit~\cite{John1994,realself1,realself2}) if $N=4$. This can be understood from the energy bands in Fig.~\figpanel{fig1}{b} and from Eq.~(\ref{decayrate}): for $\phi=(2n+1/2)\pi$ and $\delta=2\beta=2J$, the phase accumulation of the left-moving (right-moving) field traveling between the two coupling points becomes $\theta=kN=\pi N/4$ ($\theta=-3\pi N/4$), so that the spontaneous emission of the atom is allowed ($\Gamma_{\text{eff}}\neq0$) when $N=2$ and suppressed ($\Gamma_{\text{eff}}=0$) when $N=4$. Note that in Fig.~\figpanel{fig2}{b} the fractional decay is accompanied by a pronounced loss at the initial stage (i.e., $t\lesssim N/v_{g}$). This is due to the nonnegligible retardation effect which arises from not only a larger separation $N$ but also a smaller group velocity $v_{g}$ compared with the case in Fig.~\figpanel{fig2}{a}. 

More interestingly, Fig.~\figpanel{fig2}{c} shows an undamped excitation oscillation that is independent of $N$. This is because the sawtooth lattice has a flat band in this case, coupled to which the emitted photon should be strongly localized at the two coupling points. The excitation oscillates persistently between the atom and the two coupled lattice sites ($b_{0}$ and $b_{N}$), akin to the Rabi oscillation in cavity quantum electrodynamics. We have checked that the giant atom can mediate an exchange interaction between the two coupled sites in this case, with the strength of the effective interaction dependent on the detuning $\delta$.

We also plot in Figs.~\figpanel{fig2}{d}-\figpanel{fig2}{f} the time evolutions of the lattice probability distribution $P_{m}(t)=|c_{m}(t)|^{2}$ for $N=2$ and different values of $\phi$ (the results for $N=4$ can be immediately inferred from those for $N=2$). As shown in Figs.~\figpanel{fig2}{d} and \figpanel{fig2}{e}, when the giant atom is dissipationless (or shows a fractional decay), there is only a very small amount of excitation in the lattice which is mainly confined between the two coupling points; otherwise one can observe a symmetric excitation diffusion with a narrow interference pattern between the two coupling points. In the case with a flat band, as shown in Fig.~\figpanel{fig2}{f}, the excitation cannot spread in the lattice (the group velocity is always zero if $J=\beta$ and $\phi=\pi$, regardless of the wave vector $k$) and thus it is transferred back and forth between the atom and the two coupled sites.

In fact, the above results can also be analytically illustrated by calculating the \emph{self-energy} of the giant atom, which can be given by~\cite{AFKstructured}
\begin{eqnarray}
\Sigma_{e}(z)&=&\frac{g^{2}}{2\pi}\int dk\frac{\left(1+e^{-ikN}\right)\left(1+e^{ikN}\right)}{z-\omega_{k}}\nonumber\\
&=&\mp\frac{2g^{2}}{\sqrt{(z-2J)^{2}-8J^{2}(1+\cos{\phi})}}\left(1+y_{\pm}^{N}\right),
\label{selfE}
\end{eqnarray}
where $y_{\pm}=[z-2J\pm\sqrt{(z-2J)^{2}-8J^{2}(1+\text{cos}\phi)}]/2J[1+\text{exp}(i\phi)]$ and we have assumed $\beta=J$ as that in Fig.~\ref{fig2}. The Lamb shift and effective decay of the giant atom are described by the real and imaginary parts of $\Sigma_{e}(z)$, respectively. Under the Weisskopf-Wigner approximation (i.e., with weak atom-lattice interaction), one can predict most of the results in Fig.~\ref{fig2} with the self-energy $\Sigma_{e}(\delta+i0^{+})$ close to the real axis~\cite{realself1,realself2}. For example, in the case of $\delta=2J$ where the atom is resonant with the effective band center, we have 
\begin{equation}
\Sigma_{e}(2J+i0^{+})\approx\pm\frac{ig^{2}}{J\sqrt{2(1+\cos{\phi})}}\left[1+\left(\pm\frac{i\sqrt{2(1+\cos{\phi})}}{1+e^{i\phi}}\right)^{N}\right],
\label{simself}
\end{equation} 
which is zero (i.e., fractional decay) if $\phi=0$ and $N=2$ or $\phi=\pi/2$ and $N=4$, and shows a non-zero imaginary part (i.e., complete decay) if $\phi=0$ and $N=4$ or $\phi=\pi/2$ and $N=2$. Note that this formalism does not work when $\phi=(2n+1)\pi$ since in this case the energy band collapses into a single energy level.

\section{Chiral spontaneous emission}

\begin{figure}[ptb]
\centering
\includegraphics[scale=0.5]{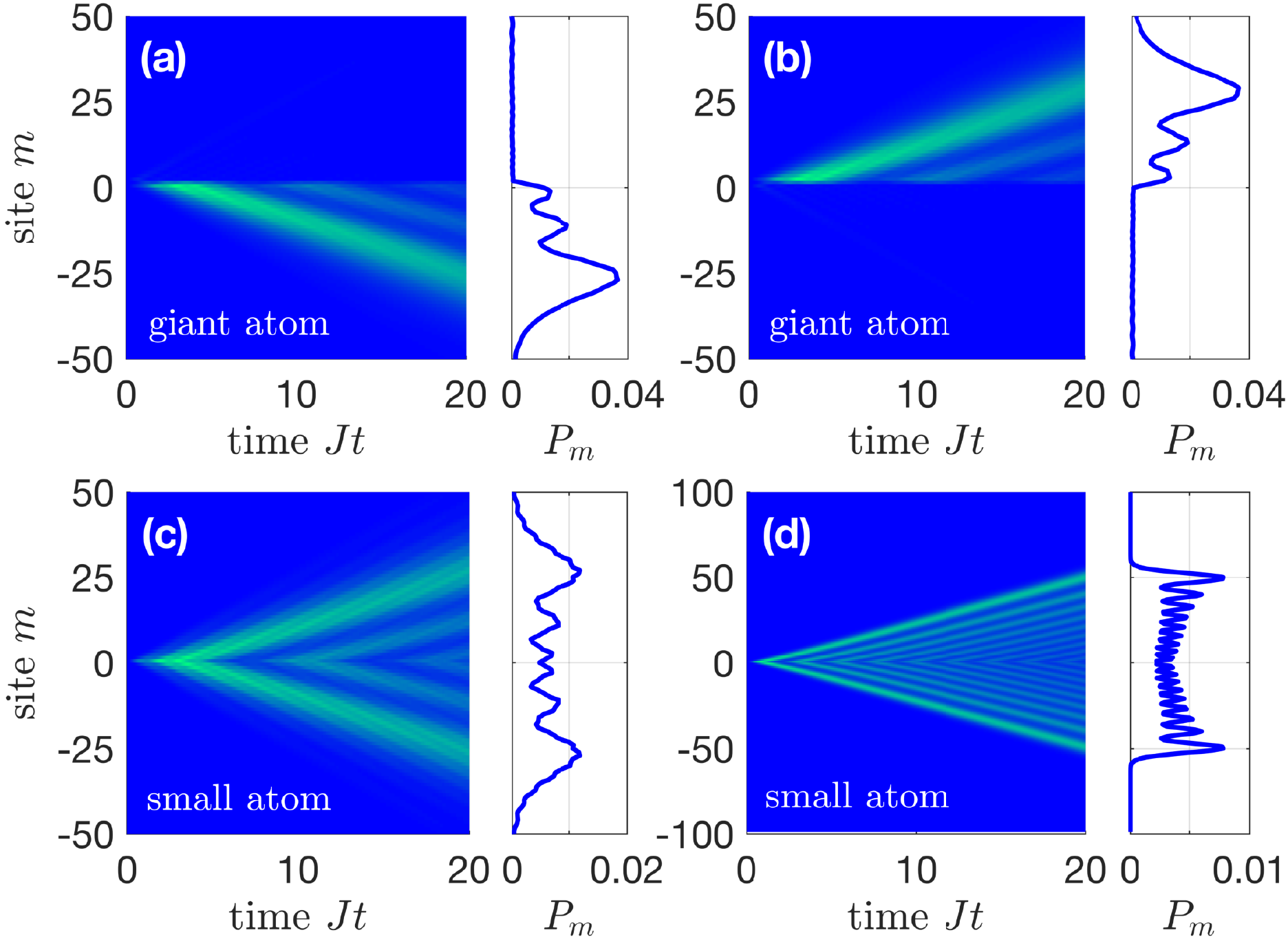}
\caption{Dynamics of lattice probability distribution $P_{m}(t)$ with (a) a giant atom of $\delta=4J$, (b) a giant atom of $\delta=0$, (c) a small atom of $\delta=4J$, and (d) a small atom of $\delta=2J$. The auxiliary charts depict the lattice distribution profiles $P_{m}(t=20/J)$. Other parameters are $g/J=0.2$, $\beta/J=1$, $\phi=\pi/2$, $N=2$, and $m_{\text{tot}}=200$.}
\label{fig3}
\end{figure}

Another intriguing feature of our model is that it enables the realization of chiral quantum optics~\cite{chiralZoller,chiralNN}. To show that, we plot in Figs.~\figpanel{fig3}{a} and \figpanel{fig3}{b} the decay dynamics of the giant atom in the case of $\phi=(2n+1/2)\pi$ and $\delta\neq2\beta$ (the atomic frequency still falls within the effective energy band). Interestingly, the giant atom exhibits a directional spontaneous emission (i.e., the atom only emits towards the left or right) if the detuning is pinned at $\delta=2(\beta\pm J)$ exactly. This phenomenon can be understood again from Fig.~\figpanel{fig1}{b}: for $\phi=(2n+1/2)\pi$, $\delta=2(\beta\pm J)$, and $N=2$, the phase accumulation towards one direction is \emph{trivial} ($\theta=0$) so that the atomic spontaneous emission is allowed, while that towards the other direction $\theta=\pm\pi$ leads to inhibited spontaneous emission due to the destructive interference. Note that the chiral spontaneous emission cannot be achieved with small atoms (i.e., atoms that are only coupled to one site of $B$) due to the absence of such directional-dependent giant-atom interference effects. As shown in Figs.~\figpanel{fig3}{c} and \figpanel{fig3}{d}, even if the effective lattice has a $k$-space asymmetric energy band, the spontaneous emission of a small atom is always symmetric (achiral) regardless of the value of $\delta$. Changing $\delta$ within the energy band only alters the group velocity of the emitted photon.

\begin{figure}[ptb]
\centering
\includegraphics[scale=0.5]{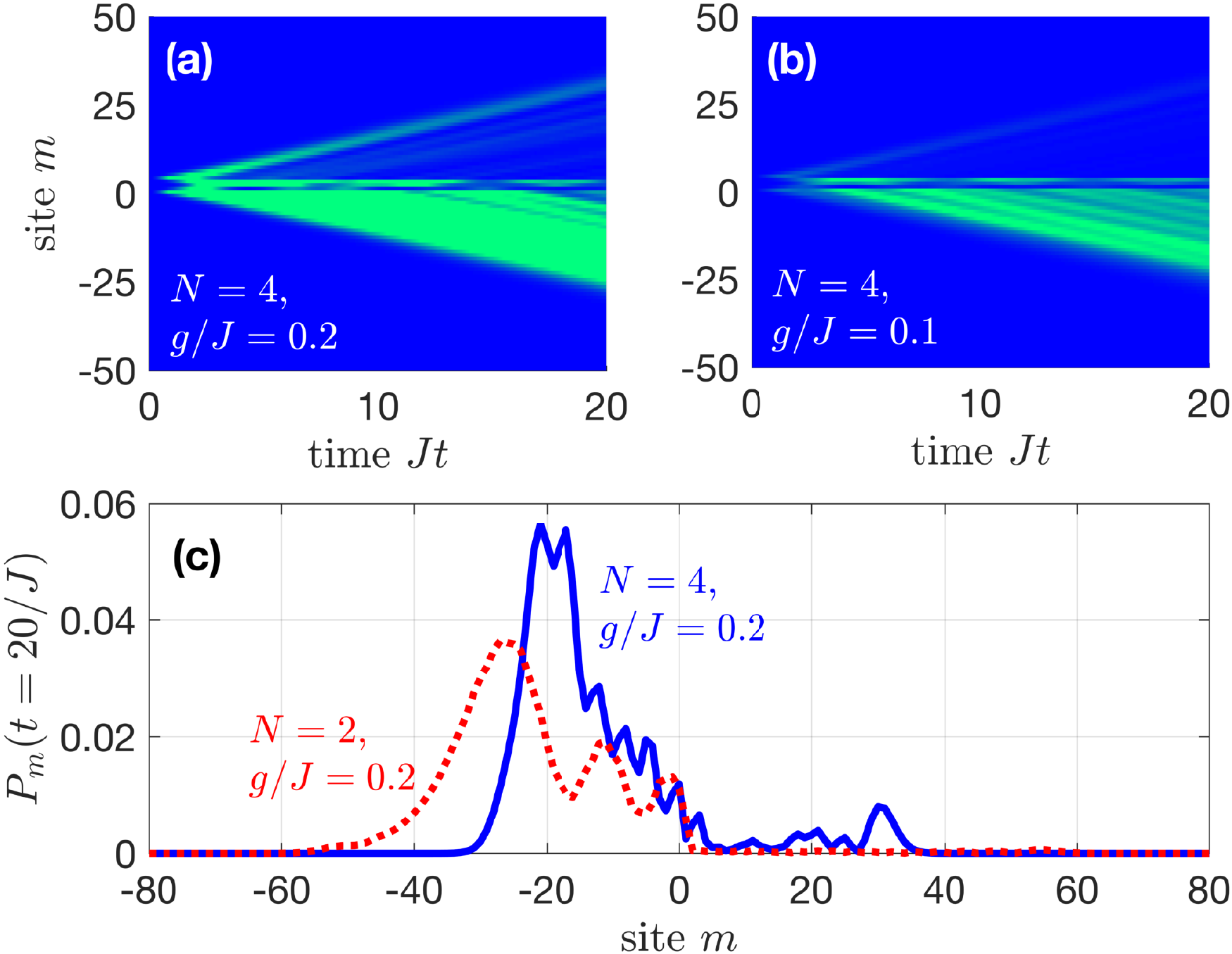}
\caption{(a, b) Dynamics of lattice probability distribution $P_{m}(t)$ for $N=4$ with (a) $g/J=0.2$ and (b) $g/J=0.1$. (c) Lattice distribution profiles $P_{m}(t=20/J)$ of the directional-emission cases for different values of $N$. In all panels we assume $\delta=4J$ and $\phi=\pi/2$ for the case of $N=2$ and $\delta=2.586J$ and $\phi=3\pi/4$ for the case of $N=4$. Other parameters are $\beta/J=1$ and $m_{\text{tot}}=200$.}
\label{fig4}
\end{figure}

The directional spontaneous emission can also be achieved for other values of $N$. As an example, we plot in Fig.~\figpanel{fig4}{a} the time evolution of $P_{m}(t)$ for $N=4$ and $\phi=3\pi/4$. In order for the giant atom to emit only towards the left (right), the detuning should be chosen as $\delta=2.586J$ ($\delta=1.414J$) [corresponding to wave vectors $k=n\pi$ and $k=(n-3/4)\pi$; see the red dotted and dashed lines in Fig.~\figpanel{fig1}{b}]. However, the directional emission is not perfect in this case due to the strong retardation effect (which arises from a larger coupling separation $N$ and a smaller group velocity $v_{g}$ as mentioned above). The retardation effect can be mitigated by assuming a smaller atom-lattice coupling strength $g$ (leading to a larger atomic relaxation time), as shown in Fig.~\figpanel{fig4}{b}. Moreover, we compare in Fig.~\figpanel{fig4}{c} the lattice distribution profiles $P_{m}(t=20/J)$ of the two directional-emission cases for $N=2$ and $N=4$. It shows that the emitted photon spreads with a smaller group velocity for $N=4$, as also predicted by the energy bands in Fig.~\figpanel{fig2}{b}. We thus conclude that giant atoms with different transition frequencies can be designed to emit photons towards different directions and with different group velocities. This can be used for spatial separations of different frequency components emitted from, e.g., a single multilevel giant atom or multiple two-level giant atoms with different frequencies.

An immediate application of such chirality is the realization of nonreciprocal delayed light, i.e., photons with certain frequencies can be retarded in one direction but not in the other. We provide in Appendix~\ref{appb} a proof-of-principle numerical simulation, which shows that a single-photon Gaussian wave packet can be absorbed by the giant atom when traveling in the direction of the chiral spontaneous emission, while the atom becomes almost ``transparent'' when a wave packet with the same frequency propagates in the opposite direction. Moreover, we reveal in Appendix~\ref{appc} that the chiral atom-field interaction \emph{cannot} result in nonreciprocal single-photon scatterings if the intrinsic dissipation of the atomic excited state is neglected. This is consistent with the results in Refs.~\cite{ChenCP,DLprr1}, where nonreciprocal scattering phenomena are allowed due to the chiral atom-field interactions \emph{as well as} the extra decay channels (e.g., dissipations into another waveguide or an additional inelastic scattering process).

\begin{figure}[ptb]
\centering
\includegraphics[scale=0.47]{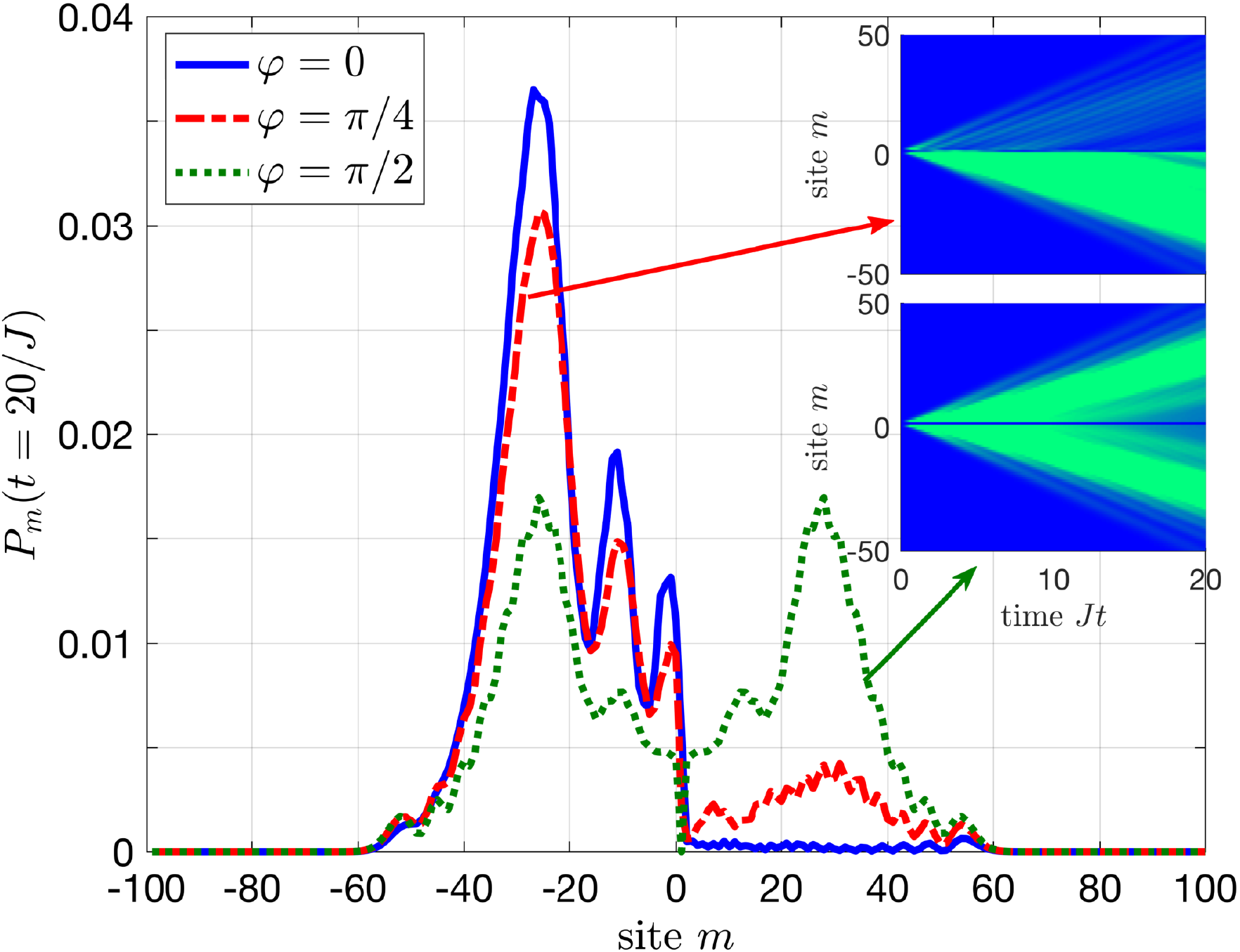}
\caption{Lattice distribution profiles $P_{m}(t=20/J)$ for different values of additional phase difference $\varphi$. The insets depict the corresponding dynamics of $P_{m}(t)$. Other parameters are $g/J=0.2$, $\beta/J=1$, $\phi=\pi/2$, $N=2$, $\delta=4J$, and $m_{\text{tot}}=200$.}
\label{fig5}
\end{figure}

Recently, it has been shown that the overall interaction between a giant atom and the waveguide field can be chiral if an additional phase difference $\varphi$ between different atom-field coupling points is introduced~\cite{DLprl,WXchiral2,ChenCP,ZollerAB}. There is a competition, however, between these two chiral mechanisms which arise from the additional phase difference $\varphi$ and the synthetic magnetic flux $\phi$, respectively. As shown in Fig.~\ref{fig5}, the lattice distribution profile $P_{m}(t=20/J)$ changes from a nearly perfect chiral shape to a symmetric one as $\varphi$ varies from $2n\pi$ to $(2n+1/2)\pi$. Physically, this is because the phase difference $\varphi$ imprints a momentum kick on the field (the phase accumulation between the two coupling points is given by $\phi'=kN+\theta$ in this case; see Appendix~\ref{appd} for more details) and thus modifies the propagation behavior of the field in the lattice. When $\theta=\phi=\pi/2$, $N=2$, and $\delta=2(\beta+J)=4J$, the left- and right-moving phase accumulations become $\phi'=\pm\pi/2$, respectively, such that the symmetric spontaneous emission reappears. Although both mechanisms allow for chiral spontaneous emission, the synthetic magnetic field breaks the time-reversal symmetry of the whole bath and thus provides a \emph{global} control when more giant atoms are involved. On the other hand, the additional phase difference can work as a \emph{local} modification of the synthetic magnetism, provided that the flux is not exactly uniform in practice.

Finally, we would like to briefly discuss the validity of the effective single-band lattice model. As mentioned above, the adiabatic elimination of sublattice $A$ is safe as long as $|\Delta|\gg\lambda$ and if the atom does not interact with $A$. In Appendix~\ref{appe}, we provide a numerical verification by comparing the result in Fig.~\figpanel{fig3}{a} with that obtained from the exact Hamiltonian [described by Eqs.~(\ref{eom1})-(\ref{eom3})], which show good agreement for $\Delta/\lambda=10$ and $\lambda/J=10$ (thus $\beta=J$ as assumed in Fig.~\ref{fig3}). Although in practice the unavoidable intrinsic loss of sublattice $A$ can introduce a finite anti-Hermitian part to the effective hopping term~\cite{LonghiBi}, we demonstrate in Appendix~\ref{appe} that our results are resistant to the intrinsic loss if the loss rate is much smaller than $|\Delta|$.

\section{Conclusions and outlook}

In summary, we have considered a sawtooth lattice (serving as a structured bath) where each triangle plaquette is threaded by a synthetic magnetic flux and one sublattice is far detuned from the other. We have also considered a two-level giant atom that is only coupled to the sublattice with nearest-neighbor couplings. When studying the decay dynamics of the giant atom, the sawtooth lattice is equivalent to a single-band structured bath with flux-controlled energy band. The synthetic magnetic flux can significantly affect the decay dynamics of the giant atom and the propagation of the emitted photon in the lattice. By tuning the flux and the transition frequency of the atom, one can realize chiral spontaneous emission and, as a result, nonreciprocal delayed light. This chirality arises from the direction-dependent giant-atom interference effects, therefore it is unattainable with single small atoms. In particular, the emitted photon can be guided towards different directions and propagate with different group velocities, depending on the atomic transition frequency, the synthetic magnetic flux, as well as the separation of the coupling points. Moreover, we have studied the competition effect between the synthetic magnetic flux and the additional phase difference between the two atom-lattice coupling points.

The proposal in this paper can bring some immediate extensions. In the case that the intrinsic loss of sublattice $A$ is strong enough, the effective lattice features \emph{non-Hermitian} complex couplings [see Eq.~(\ref{NHhopping}) in Appendix~\ref{appe}] whose imaginary parts are responsible for some exotic phenomena, such as non-Hermiticity robust transport~\cite{LonghiBi} and light trapping~\cite{DLtrap}. It is tempting to study the interplay between giant-atom interference effects and such non-Hermiticity. One can also investigate photon-atom bound states and decoherence-free interactions by coupling giant atoms to such flux-controlled structured baths, which may find applications for constructing exotic spin Hamiltonians and quantum many-body systems~\cite{WXchiral1,NoriGA,braided,FCdeco,sawtoothAGT,FCoptica}.

\appendix
\section{Hamiltonian of the effective single-band lattice}\label{appa}

In the single-excitation subspace with the eigenstate
\begin{equation}
|\psi(t)\rangle=\sum_{m}\left[c_{a,m}(t)a_{m}^{\dag}+c_{b,m}(t)b_{m}^{\dag}\right]|g\rangle\otimes|V\rangle+c_{e}(t)\sigma_{+}|g\rangle\otimes|V\rangle,
\label{eigenstate}
\end{equation}
where $c_{a,m}(t)$ and $c_{b,m}(t)$ are the excitation probability amplitudes of the $m$th sites of sublattices $A$ and $B$, respectively, and $c_{e}(t)$ is the probability amplitude of the giant atom in its excited state, the dynamical equations of the amplitudes can be obtained as
\begin{eqnarray}
i\dot{c}_{a,m}&=&-\Delta c_{a,m}+\lambda\left(c_{b,m}+e^{i\phi}c_{b,m+1}\right), \label{eom1}\\
i\dot{c}_{b,m}&=&gc_{e}\left(\delta_{m,0}+\delta_{m,N}\right)+J(c_{b,m+1}+c_{b,m-1})\nonumber\\
&&+\lambda\left(c_{a,m}+e^{-i\phi}c_{a,m-1}\right), \label{eom2}\\
i\dot{c}_{e}&=&\delta c_{e}+g\left(c_{b,0}+c_{b,N}\right), \label{eom3}
\end{eqnarray}
where we have assumed that the giant atom is coupled to the $0$th and $N$th sites of sublattice $B$. If $\Delta$ is large enough compared with $\lambda$~\cite{LonghiBi}, one can adiabatically eliminate sublattice $A$ with $\dot{c}_{a,m}\simeq 0$ such that Eqs.~(\ref{eom2}) and (\ref{eom3}) become
\begin{eqnarray}
i\dot{c}_{m}&=&2\beta c_{m}+\xi c_{m+1}+\xi^{*} c_{m-1}+gc_{e}(\delta_{m,0}+\delta_{m,N}), \label{eeom1}\\
i\dot{c}_{e}&=&\delta c_{e}+g\left(c_{0}+c_{N}\right), \label{eeom2}
\end{eqnarray}
where $\beta=\lambda^{2}/\Delta$, $\xi=J+\beta\text{exp}(i\phi)$, and we have omitted the subscript $b$ in $c_{b,m}$ for simplicity. Equations~(\ref{eeom1}) and (\ref{eeom2}) are identical with  Eqs.~(\ref{de1}) and (\ref{de2}), implying that the effective model after adiabatic elimination can be described in the single-excitation subspace by Hamiltonian (\ref{effectH}) in the main text.


\section{Nonreciprocal delayed light}\label{appb}

\begin{figure}[ptb]
\centering
\renewcommand {\thefigure} {A1}
\includegraphics[scale=0.7]{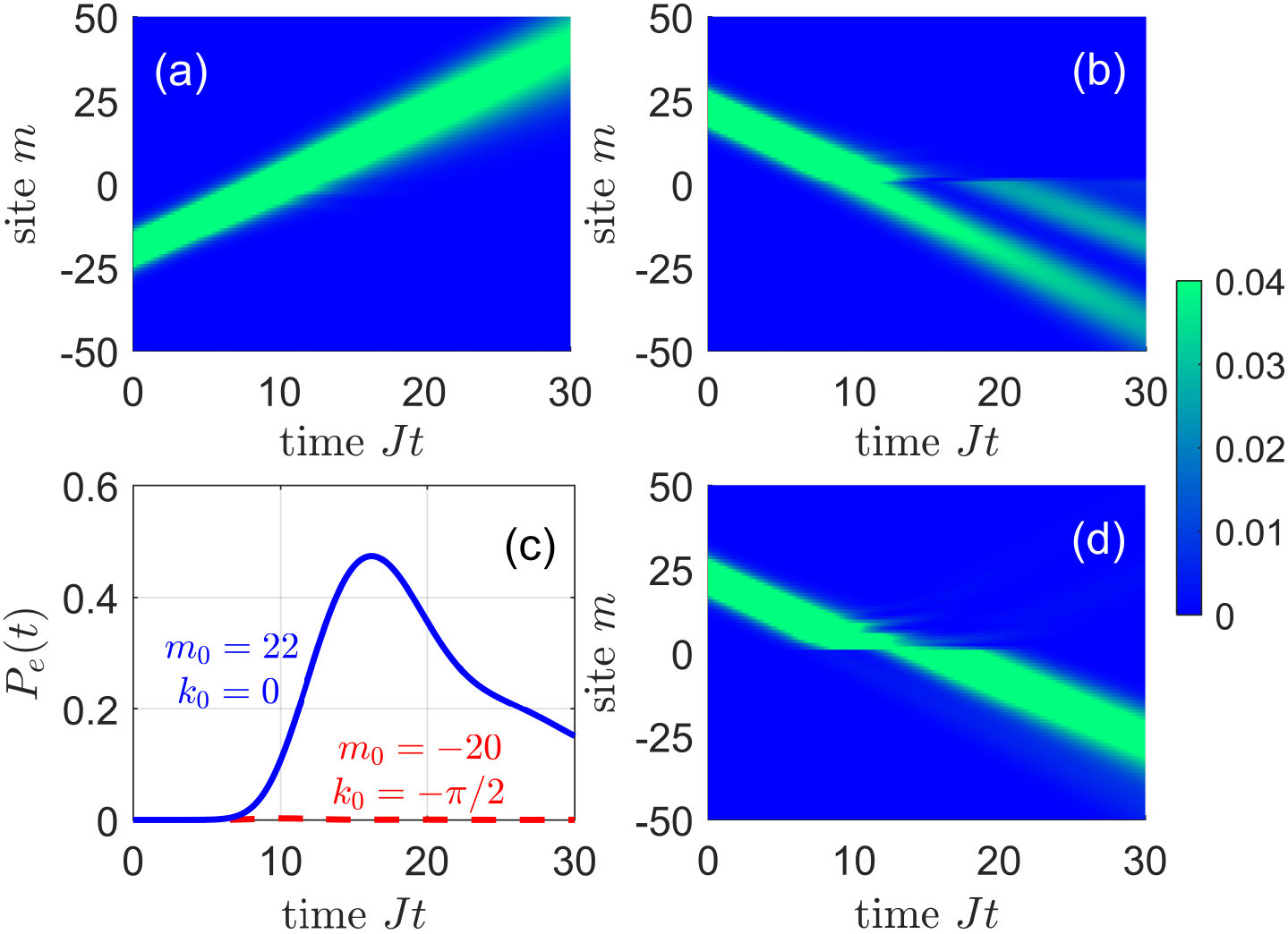}
\caption{(a, b) Dynamics of lattice probability distribution $P_{m}(t)$ for (a) $m_{0}=-20$ and $k_{0}=-\pi/2$ and (b) $m_{0}=22$ and $k_{0}=0$ with a single giant atom. (c) Dynamics of excitation probability $P_{e}(t)$ of the single giant atom for the two incident waves in panels (a) and (b). (d) Dynamics of lattice probability distribution $P_{m}(t)$ for $m_{0}=22$ and $k_{0}=0$ with $8$ identical giant atoms. Other parameters are $g/J=0.2$, $\delta=4J$, $\phi=\pi/2$, $N=2$, $w=5$, and $m_{\text{tot}}=200$.}
\label{figa1}
\end{figure}

To demonstrate the nonreciprocal light delay effect, we consider in this appendix Gaussian-shape incident waves
\begin{equation}
c_{m}(t=0)=A\text{exp}\left[-\frac{(m-m_{0})^{2}}{2w^{2}}+ik_{0}m\right],
\label{Gaussian}
\end{equation}
where $m_{0}$ is the initial position of the wave center, $w$ is the spatial width, $k_{0}$ is the carrier wave vector, and $A=\pi^{-1/4}w^{-1/2}$ is the normalization coefficient.

As shown in Fig.~\figpanel{fig3}{a}, when $\phi=\pi/2$, $N=2$, and $\delta=4J$, the giant atom only emits towards the left since the right-moving emitted field undergoes a destructive interference. This also implies that a right-moving resonant photon (with wave vector $k=-\pi/2$) cannot excite the giant atom in this case. In Figs.~\figpanel{figa1}{a} and \figpanel{figa1}{b}, we plot the time evolutions of the lattice probability distribution $P_{m}(t)$ for two Gaussian-shape incident waves with the same frequency, i.e., (i) $m_{0}=-20$ and $k_{0}=-\pi/2$ (a right-moving wave) and (ii) $m_{0}=22$ and $k_{0}=0$ (a left-moving wave), respectively. As expected, the right-moving wave cannot ``feels'' the giant atom and thus passes through the two coupling points directly, whereas a portion of the left-moving wave is retarded due to the atomic absorption [as shown in Fig.~\figpanel{figa1}{c}, this portion is absorbed by the atom and then re-emitted in the same direction]. Note that the light delay effect can be further enhanced if more such giant atoms are involved. As shown in Fig.~\figpanel{figa1}{d}, we consider a multi-atom situation where $8$ identical two-level giant atoms are coupled to the same lattice sites (i.e., $b_{0}$ and $b_{2}$). In this case, the incident wave can hardly pass through the atoms directly and is greatly absorbed by the atoms (and thus retarded).    

We would like to point out that the (nonreciprocal) delayed light here is based on the chiral atom-field interaction. This is conceptually different from the slow-light effects based on electromagnetically-induced transparency~\cite{EIT1,EIT2,EIT3}, coherent population oscillations~\cite{CPO1,CPO2}, and optomechanically-induced transparency~\cite{OMIT1,OMIT2,OMIT3}, which feature dramatic reduction in the group velocity of light due to the sharp normal dispersions in the transparency windows. To achieve nonreciprocal phenomena in these cases, one has to employ other effects such as the synthetic magnetism~\cite{OMnon1,OMnon2,OMnon3} and Sagnac effect~\cite{Sagnac1,Sagnac2} in optomechanical systems and the random thermal motions in hot atoms~\cite{hot1,hot2}. Moreover, in contrast to the nonreciprocal delayed light schemes based on stimulated Brillouin scatterings~\cite{SBS1,SBS2,SBS3}, where two counterpropagating optical fields are coupled to a traveling acoustic wave in the strong-light regime, our scheme works at the single-photon level and does not require a rapid variation of the refractive index of the background material.

\section{Single-photon scattering spectra}\label{appc}

In Appendix~\ref{appb} we have demonstrated the nonreciprocal light delay effect which arises from the chiral atom-waveguide interaction. In this appendix, we would like to reveal that the single-photon scattering in our model is however \emph{reciprocal} despite this chirality, unless other decay channels of the atom are introduced.

\begin{figure}[ptb]
\centering
\renewcommand {\thefigure} {A2}
\includegraphics[scale=0.52]{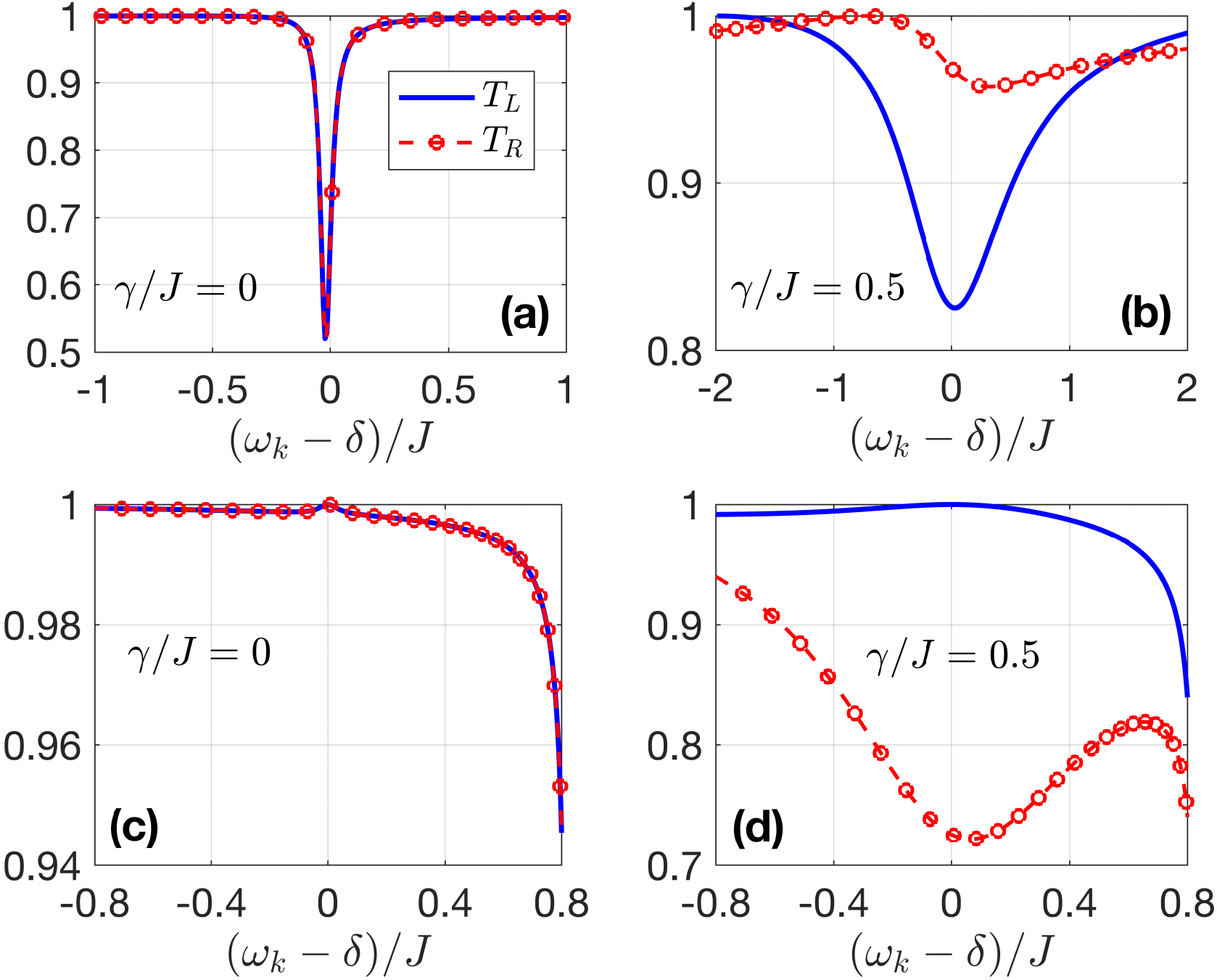}
\caption{Single-photon transmission rates $T_{L}$ and $T_{R}$ versus detuning $\omega_{k}-\delta$ for (a, b) $\delta/J=2$ and $N=3$ and (c, d) $\delta/J=4$ and $N=2$. We assume $\gamma=0$ in panels (a) and (c) and $\gamma/J=0.5$ in panels (b) and (d). Other parameters are $g/J=0.2$ and $\phi=\pi/2$.}
\label{figa2}
\end{figure}

Recalling the dynamical equations~(\ref{de1}) and (\ref{de2}), the scattering probabilities of a single-photon incident wave (with energy $\omega_{k}$) can be calculated by assuming the solutions as $c_{m}(t)=\tilde{c}_{m}\text{exp}(-i\omega_{k}t)$ and $c_{e}(t)=\tilde{c}_{e}\text{exp}(-i\omega_{k}t)$, with 
\begin{equation}
\tilde{c}_{m}=\left\{
\begin{aligned}
&e^{ikm}+r_{L}e^{ik'm}& \quad m<0, \\
&f_{L}e^{ikm}+f_{L}'e^{ik'm}& \quad 0\leqslant m\leqslant N, \\
&t_{L}e^{ikm}& \quad m>N
\end{aligned}
\right.
\label{leftin}
\end{equation}
for a left-incident wave, and
\begin{equation}
\tilde{c}_{m}=\left\{
\begin{aligned}
&t_{R}e^{ik'm}& \quad m<0, \\
&f_{R}'e^{ik'm}+f_{R}e^{ikm}& \quad 0\leqslant m\leqslant N, \\
&e^{ik'm}+r_{R}e^{ikm}& \quad m>N
\end{aligned}
\right.
\label{rightin}
\end{equation}
for a right-incident wave. In Eqs.~(\ref{leftin}) and (\ref{rightin}), $k$ and $k'$ represent the wave vectors of the right-moving and left-moving fields, respectively; $t_{L}$ and $t_{R}$ ($r_{L}$ and $r_{R}$) are the transmission (reflection) amplitudes of the left-incident and right-incident waves, respectively; $f_{L}$ and $f_{R}$ ($f_{L}'$ and $f_{R}'$) describe the right-moving (left-moving) fields between the two coupling points in the two incident cases. Note that $k$ and $k'$ satisfy $k+k'+\phi=0$ which reduces to the common relation $k=-k'$ in the time-reversal-unbroken case (i.e., $\phi=0$), as shown in Fig.~\figpanel{fig1}{b}. Using the solutions above, one can obtain the stationary-state equations
\begin{eqnarray}
&&(\omega_{k}-2\beta)\tilde{c}_{-1} = \xi\tilde{c}_{0}+\xi^{*}\tilde{c}_{-2}, \label{sse1}\\
&&(\omega_{k}-2\beta)\tilde{c}_{0} = \xi\tilde{c}_{1}+\xi^{*}\tilde{c}_{-1}+g\tilde{c}_{e}, \label{sse2}\\
&&(\omega_{k}-2\beta)\tilde{c}_{N} = \xi\tilde{c}_{N+1}+\xi^{*}\tilde{c}_{N-1}+g\tilde{c}_{e}, \label{sse3}\\
&&(\omega_{k}-2\beta)\tilde{c}_{N+1} = \xi\tilde{c}_{N+2}+\xi^{*}\tilde{c}_{N}, \label{sse4}\\
&&(\omega_{k}-\delta)\tilde{c}_{e} = g(\tilde{c}_{0}+\tilde{c}_{N}), \label{sse5}
\end{eqnarray}
and calculate the scattering amplitudes of the two incident cases.

We plot in Fig.~\ref{figa2} the single-photon transmission rates $T_{L}=|t_{L}|^{2}$ (for a left-incident wave) and $T_{R}=|t_{R}|^{2}$ (for a right-incident wave). One can see from Figs.~\figpanel{figa2}{a} and \figpanel{figa2}{c} that the single-photon transmissions are always reciprocal (i.e., $T_{L}\equiv T_{R}$). This is consistent with the results in Refs.~\cite{ChenCP,DLprr1}, where chiral atom-field interactions cannot result in nonreciprocal single-photon scatterings if other decay channels of the atomic excited state are neglected. Indeed, as shown in Figs.~\figpanel{figa2}{b} and \figpanel{figa2}{d}, the transmissions become nonreciprocal if we consider a finite intrinsic decay rate $\gamma$ for the giant atom, i.e., $-\delta\rightarrow-\delta+i\gamma$ in Eq.~(\ref{sse5}). As a result, nonreciprocal single-photon scattering behaviors arise from the combination of chiral atom-field interactions and external decay channels of the scattering center (i.e., the atom).

\section{Momentum kick induced by the additional phase difference}\label{appd}

If we introduce an additional phase difference $\varphi$ through the atom-lattice interaction term~\cite{DLprl,ZollerAB}
\begin{equation}
H_{\text{int}}=g\sigma_{+}[b_{0}+\text{exp}(i\varphi)b_{N}]+\text{H.c.}
\label{VincludeVarphi}
\end{equation}
and perform the transformation
\begin{equation}
b_{k}=\frac{1}{\sqrt{2\pi}}\sum_{m}b_{m}e^{-ikm},
\label{transform}
\end{equation}
Eq.~(\ref{effectH}) becomes
\begin{eqnarray}
H&=&H_{0}+H_{\text{int}}', \label{kick1}\\
H_{0}&=&\delta\sigma_{+}\sigma_{-}+\int dk\omega_{k}b_{k}^{\dag}b_{k}, \label{kick2}\\
H_{\text{int}}'&=&G\int dk\left[\left(1+e^{-i(kN+\varphi)}\right)\sigma_{-}b_{k}^{\dag}+\text{H.c.}\right] \label{kick3}
\end{eqnarray}
with $\omega_{k}$ given in Eq.~(\ref{dispersion}) and $G=g/\sqrt{2\pi}$. Equation~(\ref{kick3}) shows a standard atom-field interaction but with an engineered $k$-dependent coupling rate. The additional phase difference $\varphi$ renders the coupling asymmetric with respect to $k=0$ and thus imprints a momentum kick on the emitted photons towards different directions. This effect competes with the synthetic magnetism of the lattice, as shown in Fig.~\ref{fig5} in the main text.

\section{Numerical verifications with the exact Hamiltonian}\label{appe}

\begin{figure}[ptb]
\centering
\renewcommand {\thefigure} {A3}
\includegraphics[scale=0.55]{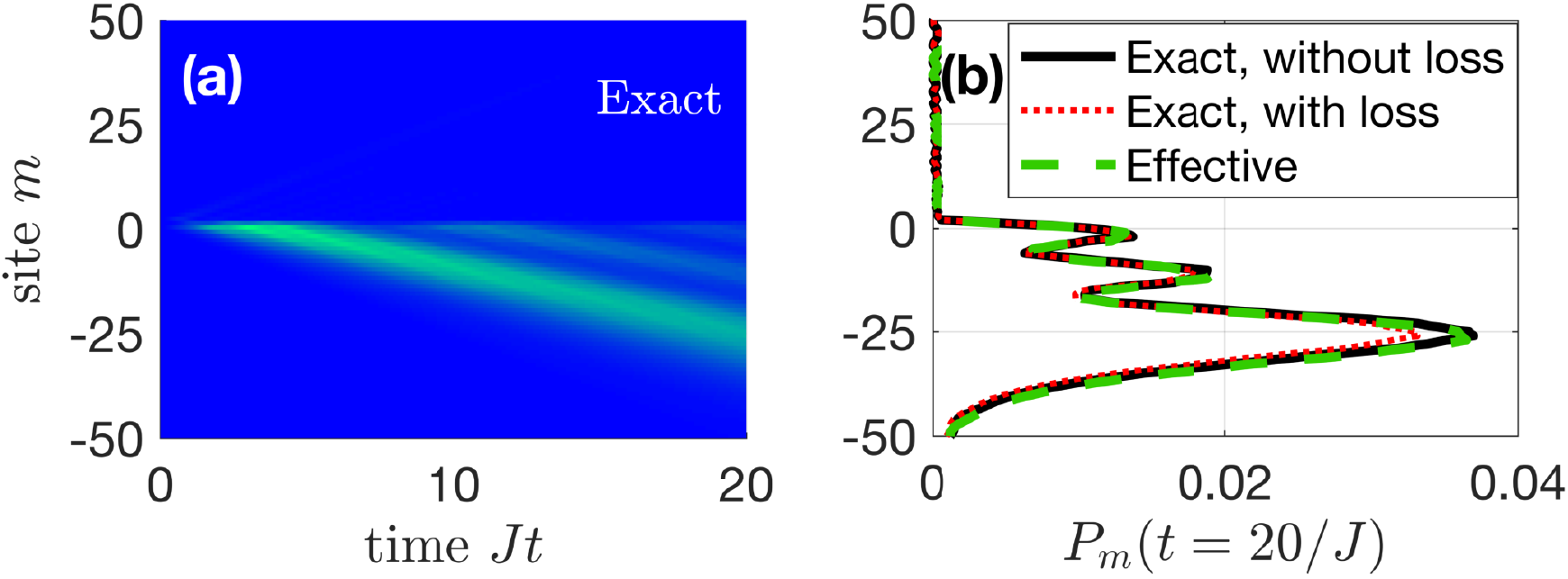}
\caption{(a) Dynamics of lattice probability distribution $P_{m}(t)$ obtained from the exact Hamiltonian [described by Eqs.~(\ref{eom1})-(\ref{eom3})]. (b) Lattice distribution profiles $P_{m}(t=20/J)$ obtained from the exact and effective Hamiltonians (with and without intrinsic loss of the $A$ sites). Other parameters are $g/J=0.2$, $\lambda/J=10$, $\Delta/J=100$, $\delta=4J$, $\phi=\pi/2$, $\kappa/J=0.2$, $N=2$, and $m_{\text{tot}}=200$.}
\label{figa3}
\end{figure}

In this appendix we numerically verify the validity of our proposal. To this end, we calculate the dynamical equations Eqs.~(\ref{eom1})-(\ref{eom3}) obtained from the exact Hamiltonian, and compare the results with those obtained from the effective Hamiltonian Eq.~(\ref{effectH}). As an example, we plot in Fig.~\figpanel{figa3}{a} the dynamics of the lattice probability distribution $P_{m}(t)$ with the parameters in Fig.~\figpanel{fig3}{a} and with $\lambda/J=10$ and $\Delta/\lambda=10$. It shows a nearly identical time evolution with that in Fig.~\figpanel{fig3}{a}. Moreover, we compare in Fig.~\figpanel{figa3}{b} the profiles of the lattice probability distribution at $t=20/J$ obtained from the exact and effective Hamiltonians, which also show good agreement with matched parameters. When taking the intrinsic loss of the $A$ sites into account [i.e., $-\Delta\rightarrow-\Delta-i\kappa$ in Eq.~(\ref{sawtoothH})], the profile shows a slight probability reduction yet its chiral shape maintains almost unchanged. Although the hopping term 
\begin{equation}
\sum_{m}\left(J+\frac{\lambda^{2}}{\Delta+i\kappa}\right)\left(e^{i\phi}b_{m}^{\dag}b_{m+1}+\text{H.c.}\right)
\label{NHhopping}
\end{equation}
of the effective lattice becomes \emph{non-Hermitian} for $\kappa\neq0$, the effect of the anti-Hermitian part can be neglected if $|\Delta|\gg\kappa$. We thus conclude that our results in this paper are resistant to weak intrinsic loss of sublattice $A$.

\section*{Acknowledgment}

This work was supported by the National Natural Science Foundation of China (under Grants No. 12074030, No. 12074061, and No. 12274107), National Key Research and Development Program of China (No. 2021YFE0193500), and the Research Funds of Hainan University [Grant No. KYQD(ZR)23010].

\section*{Data availability statement}
The data that support the findings of this study are available upon reasonable request from the authors.


\end{document}